\documentclass{article}
\usepackage[english]{babel}
\usepackage[utf8]{inputenc}
\usepackage{johd}
\usepackage{authblk}

\title{Dflow, a Python framework for constructing cloud-native 
AI-for-Science
workflows}

\author[1]{\textbf{Xinzijian Liu}}
\author[2]{\textbf{Yanbo Han}}
\author[3]{\textbf{Zhuoyuan Li}}
\author[4]{\textbf{Jiahao Fan}}
\author[5,6]{\textbf{Chengqian Zhang}}
\author[7]{\textbf{Jinzhe Zeng}}
\author[8]{\textbf{Yifan Shan}}
\author[1]{\textbf{Yannan Yuan}}
\author[9]{\textbf{Wei-Hong Xu}}
\author[10]{\textbf{Yun-Pei Liu}}
\author[1]{\textbf{Yuzhi Zhang}}
\author[3]{\textbf{Tongqi Wen}}
\author[7]{\textbf{Darrin M. York}}
\author[11,12]{\textbf{Zhicheng Zhong}}
\author[1]{\textbf{Hang Zheng}}
\author[9,10,13]{\textbf{Jun Cheng}}
\author[1,5, \thanks{\href{mailto:linfeng.zhang.zlf@gmail.com}{linfeng.zhang.zlf@gmail.com}}]{\textbf{Linfeng Zhang}}
\author[6,14, \thanks{\href{mailto:wang_han@iapcm.ac.cn}{wang\_han@iapcm.ac.cn}}]{\textbf{Han Wang}}

\affil[1]{DP Technology, Beijing 100080, China}
\affil[2]{Institute of Molecular Science and Applied Chemistry, School of Chemistry,  Xi’an Jiaotong University, Xi’an 710049, China}
\affil[3]{Department of Mechanical Engineering, The University of Hong Kong, Hong Kong, China}
\affil[4]{School of Physics, Peking University, Beijing 100871, China}
\affil[5]{AI for Science Institute, Beijing 100080, China}
\affil[6]{HEDPS, CAPT, College of Engineering, Peking University, Beijing 100871, China}
\affil[7]{Laboratory for Biomolecular Simulation Research, Center for Integrative Proteomics Research, and Department of Chemistry and Chemical Biology, Rutgers University, Piscataway, New Jersey 08854-8087, United States}
\affil[8]{Ningbo Institute of Materials Technology and Engineering, Chinese Academy of Sciences, Ningbo 315201, China}
\affil[9]{Laboratory of AI for Electrochemistry (AI4EC), IKKEM, 361005 Xiamen, China}
\affil[10]{State Key Laboratory of Physical Chemistry of Solid Surface, Collaborative Innovation Center of Chemistry for Energy Materials, College of Chemistry and Chemical Engineering, Xiamen University, Xiamen 361005, China}
\affil[11]{Department of Physics, University of Science and Technology of China, Hefei 230026, China}
\affil[12]{Suzhou Institute for Advanced Research, University of Science and Technology of China, Suzhou 215123, China}
\affil[13]{Institute of Artificial Intelligence, Xiamen University, Xiamen 361005, China}
\affil[14]{Laboratory of Computational Physics, Institute of Applied Physics and Computational Mathematics, Beijing 100094, China}

\date{}

\begin{document}
\maketitle
\begin{abstract} 
\noindent In the AI-for-science era, scientific computing scenarios such as concurrent learning and high-throughput computing demand a new generation of infrastructure that supports scalable computing resources and automated workflow management on both cloud and high-performance supercomputers. Here we introduce Dflow, an open-source Python toolkit designed for scientists to construct workflows with simple programming interfaces. It enables complex process control and task scheduling across a distributed, heterogeneous infrastructure, leveraging containers and Kubernetes for flexibility. Dflow is highly observable and can scale to thousands of concurrent nodes per workflow, enhancing the efficiency of complex scientific computing tasks. The basic unit in Dflow, known as an Operation (OP), is reusable and independent of the underlying infrastructure or context. Dozens of workflow projects have been developed based on Dflow, spanning a wide range of projects. We anticipate that the reusability of Dflow and its components will encourage more scientists to publish their workflows and OP components. These components, in turn, can be adapted and reused in various contexts, fostering greater collaboration and innovation in the scientific community.
\end{abstract}

\noindent\\

\section{Introduction}

Advancements in artificial intelligence (AI) are propelling a transformative approach to scientific discovery. This shift, characterized by substantial improvements in computational power and theoretical methodologies, paves the way for the development of sophisticated machine learning models capable of tackling complex scientific questions. 
Despite early successes, the integration of scientific data and computing with AI reveals intricate challenges. 
{For instance, the iterative closed-loop process of efficient data sampling or generation for training machine learning interatomic potentials in the framework of active learning~\cite{smith2018less} or concurrent learning~\cite{zhang2020dp}, exemplifies the complexity involved in achieving high-fidelity machine learning models for scientific applications. }

The operationalization of these computational tasks further underscores the need for a more efficient approach. The traditional manual management of tasks on supercomputers, characterized by a repetitive cycle of submission, waiting, and processing, is markedly inefficient. This inefficiency is compounded by the use of simple scripted automation, which, while a step towards simplification, often suffers from poor reusability and maintainability due to the tight coupling of operations and the absence of abstracted frameworks. Such limitations are exacerbated by the lack of visibility into the processes, making problem identification and debugging increasingly challenging.

In response to these challenges, the evolving capabilities of cloud computing, particularly its rapid scalability, present a crucial opportunity. This transition to cloud computing aims to improve both the efficiency and cost-effectiveness of research, marking a significant shift away from traditional computational methods. In this context, there's an urgent need for a new, user-friendly workflow framework. Such a framework must efficiently utilize the benefits of cloud resources and high-performance computing (HPC), effectively narrowing the divide between algorithm conception and its practical application. This strategy is essential for the seamless incorporation of advanced computational resources into scientific research.

Among existing workflow management tools are AiiDA~\cite{huber2020aiida} and Fireworks~\cite{CPE:CPE3505} from the scientific computing domain, and Airflow~\cite{harenslak2021data} from the cloud-native sphere. AiiDA prioritizes reproducibility and data provenance, which could compromise backend performance and presents a considerable learning curve; Fireworks, with its dependence on local file systems or MongoDB, lacks the flexibility for integrating with cloud storage, affecting reliability, reproducibility, and costs. These systems do not fully embrace containerization or cloud computing resources. Conversely, Airflow, despite its popularity in data engineering, does not adequately address the intricate requirements of scientific computing, such as managing recursive workflows.

In this work, we introduce Dflow, an open-source Python toolkit\footnote{open-sourced under the LGPL license at \url{https://github.com/deepmodeling/dflow}.} 
designed to address the issues mentioned above and tackle the challenges in bridging the conceptual design of algorithms to their practical implementation. 
Released in January 2022, Dflow focuses on the following key features:

{\bf (1) Process Control and Reproducibility}: Dflow integrates Argo Workflows for reliable scheduling and management of tasks. Its use of containers simplifies software setup across different platforms, enhancing reproducibility. While containers are advised to ensure consistent environments, they are optional in Dflow. The addition of Kubernetes and reliable storage further strengthens workflow stability and monitoring.

{\bf (2) Adaptability to Various Environments}: Designed for a range of computing setups, Dflow's server can run anywhere from single machines to large cloud-based Kubernetes clusters. Importantly, it supports High Performance Computing (HPC) environments, crucial for scientific research. Dflow offers flexibility in managing HPC jobs, whether through specific plugins or by integrating HPC resources directly into Kubernetes, making it versatile for any computational requirement.

\textbf{{(3) Flexible Rules and Local Debugging}}:
{Exception handling and fault tolerance policies can be set before workflow submission, enabling the process to proceed as planned even if errors occur. Dflow offers a flexible restart mechanism, allowing for the seamless integration of completed steps from previous workflows into new ones.} Dflow’s debug mode allows for local workflow execution without containers, providing a consistent experience with its main Argo workflow mode. This feature is especially useful for debugging or running tests in environments where container use may be limited, highlighting Dflow's commitment to user convenience and adaptability.

To date, Dflow has been the foundation for numerous workflow projects across a diverse array of fields, from electronic structure calculation and molecular dynamics to biological simulations, drug design, material property predictions, and automated software testing. This widespread adoption is a testament to Dflow's openness and extendibility, features that encourage innovation and customization. 
Furthermore, Dflow's ecosystem has grown to include an extensive collection of reusable operations (OPs) and advanced super operations (super OPs), exemplified by tools like FPOP\footnote{\url{https://github.com/deepmodeling/fpop}}. 
These components not only facilitate the development of complex workflows but also promote a collaborative environment where researchers and developers can share advancements and best practices. 
This collaborative aspect, driven by Dflow's open and extendible architecture, ensures that it remains at the cutting edge of workflow management, offering robust support for the evolving needs of scientific and technical communities.

\section{Design and implementation}
\begin{figure}[htb]
 \center{\includegraphics[width=0.5\textwidth]{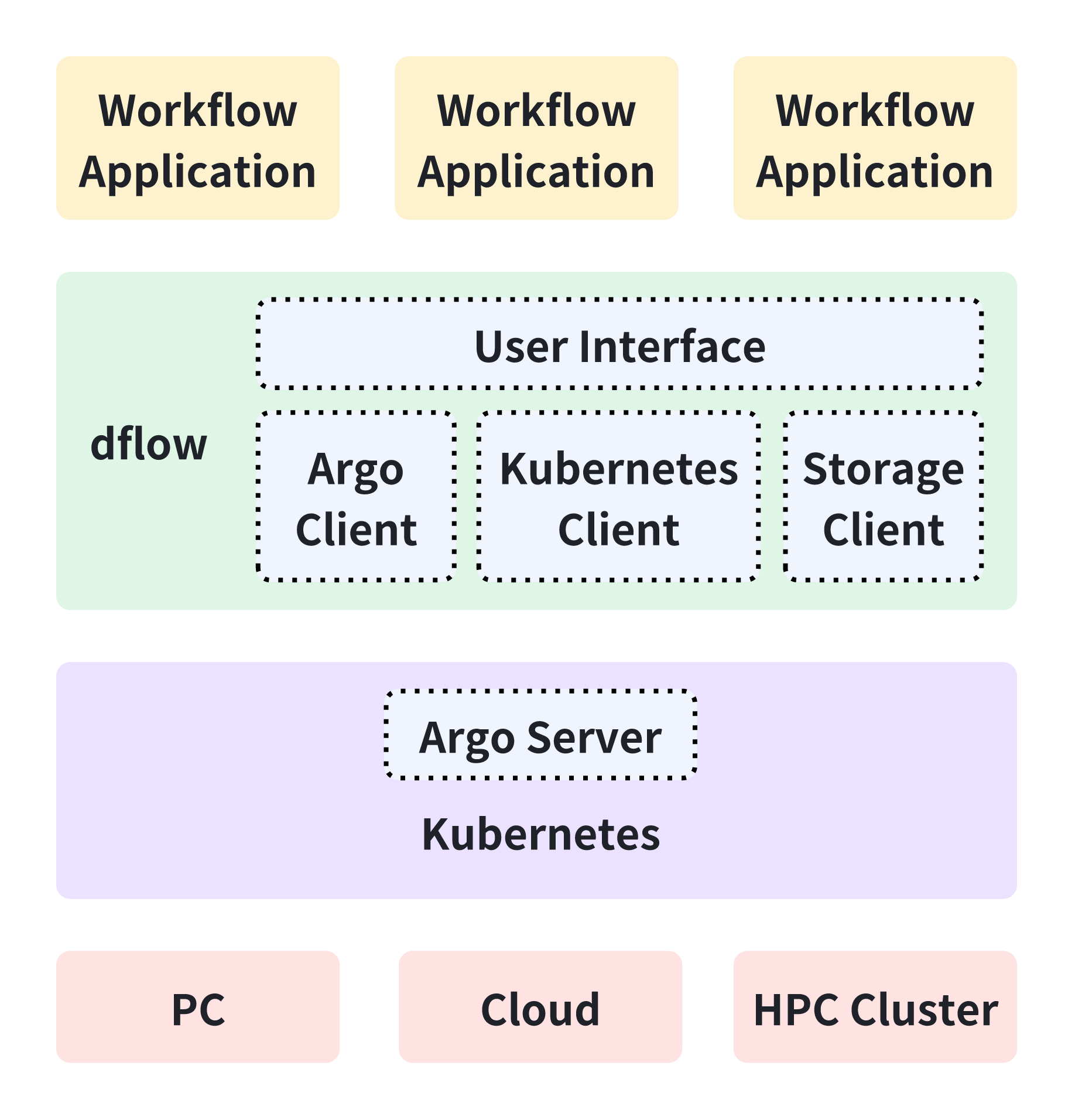}} 
 \caption{\label{dflow-architecture} A schematic graph of Dflow within the AI-for-science computing framework.} 
\end{figure}

We refer to Figure \ref{dflow-architecture} for a schematic representation of Dflow within the AI-for-science computing framework.
For those unfamiliar with the infrastructure, Argo Workflows\footnote{\url{https://argoproj.github.io/}} is an open-source, cloud-native engine for orchestrating containerized jobs on Kubernetes, supporting complex workflows and managing large-scale task execution efficiently.

As depicted in Figure \ref{dflow-architecture}, Dflow leverages Argo Workflows as its core engine, enhancing it with intuitive programming interfaces for workflow construction. This approach allows developers to concentrate on algorithmic logic and workflow orchestration, bypassing the complexities of underlying systems. Dflow further enriches user experience with a web UI and command-line tools for monitoring and managing workflows.

Key to Dflow's architecture is its use of Kubernetes and robust storage solutions, ensuring workflow observability, reproducibility, and resilience. Designed for distributed and heterogeneous environments, Dflow can operate across diverse platforms, from Minikube on a single machine to large cloud-based Kubernetes clusters. Notably, it extends its computing capabilities to high-performance computing (HPC) clusters, crucial for scientific computations. Users can dispatch HPC jobs using DPDispatcher plugins or manage HPC resources via virtual node techniques within Kubernetes, leveraging the wlm-operator for seamless integration.

In the subsequent sections, we delve deeper into the design principles and functionalities of Dflow.

\subsection{Basics}

\begin{figure}[htb]
 \center{\includegraphics[width=0.6\textwidth]{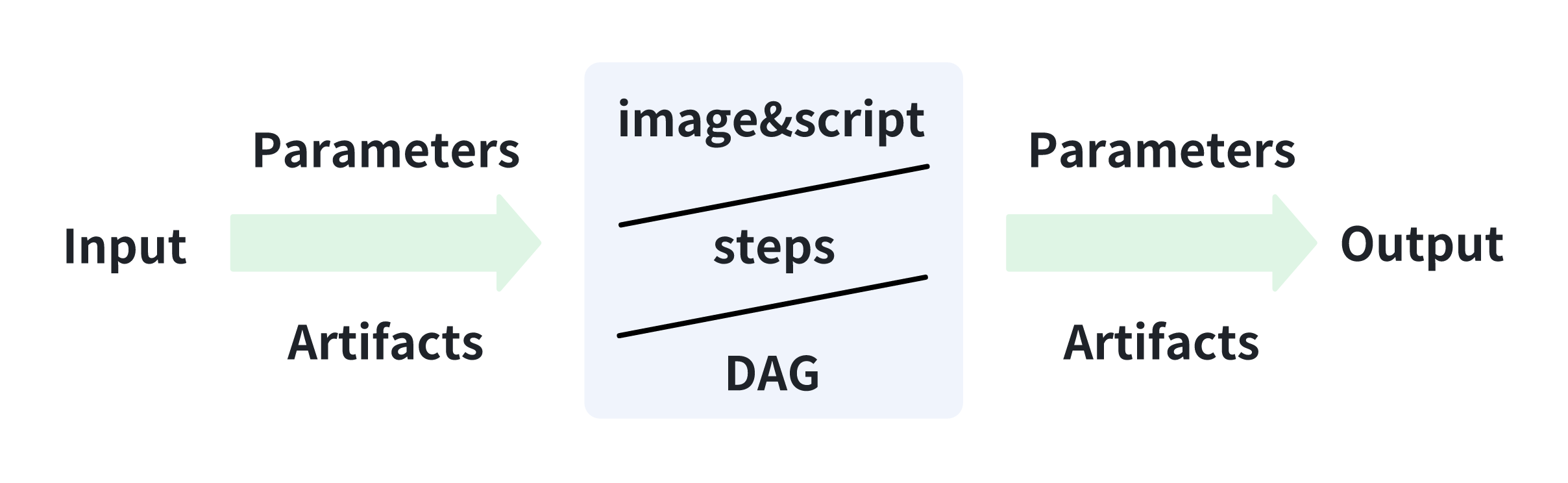}} 
 \caption{\label{op-schema} A schematic graph for OP template.} 
\end{figure}

The language of defining workflows is the central part of Dflow. Operation (abbr. OP) template serves as a fundamental building block of a workflow. It defines a particular operation to be executed given the input structure and the expected output structure. Both the input and output can be parameters and/or artifacts. Parameters and artifacts are data stored and transferred within a workflow. Parameters are saved as text which can be displayed in the UI, while artifacts are stored as files. Parameters are passed to an OP with their values, while artifacts are passed by paths. As dipicted in Figure \ref{op-schema}, an OP can be implemented by executing a script within a container, as well as through several steps or a directed acyclic graph (DAG). This decouples the workflow logic from the specific implementation inside the OP. OP extracts the shared core logic of a category of operations, and accommodates minor variations as inputs. OPs can be reused among workflows and shared among users.

The container OP template forms the foundation of all OP templates, enabling operations to be executed within containers. Two types of container OP templates are supported: \verb|ShellOPTemplate| and \verb|PythonScriptOPTemplate|. \verb|ShellOPTemplate| defines an operation by shell script and a container image where the script runs. \verb|PythonScriptOPTemplate| defines an operation by Python script and a container image.

As a more Python-native category of OP templates provided by Dflow, \verb|PythonOPTemplate| is designed to leverage Python's flexibility, allowing scientists to define operations either as classes or functions, which aligns seamlessly with the Pythonic approach to programming.

To ensure clarity and stability in operations, Dflow enforces strict type checking for Python OPs, thereby preempting ambiguity and unexpected behavior that can arise from Python's dynamic typing. For a class OP, the input and output structures of an OP are declared in the static methods \verb|get_input_sign| and \verb|get_output_sign|. Each of them returns a dictionary mapping from the name of a parameter/artifact to its type. Any serializable type, including custom objects, is an acceptable parameter type. 

The execution of the OP is defined in the \verb|execute| method, working as python native functions. For an input artifact of an OP, the OP receives a path, a list of paths or a dictionary of paths. Programmer can directly process the file(s) or directory(ies) at the path(s) without knowing the absolute path(s). For an output artifact, the OP should also returns a path, a list of paths or a dictionary of paths. Type checking is implemented before and after the \verb|execute| method.

A function OP offers a more concise approach, leveraging Python's type annotations to explicitly define input and output structures. The execution logic is encapsulated within the function body, in line with Python's syntactic norms. These function OPs are seamlessly translatable into class OPs, with type checking enforced before and after the \verb|execute| method.

Central to Dflow's workflow management is the Step, which articulates flow by instantiating OP templates with specified input values and artifact sources. This mechanism allows for both static and dynamic inputs variables, depending on the workflow's requirements.

By connecting Steps, complex workflows are constructed, with the ability to sequence them for serial execution or to run them in parallel, thereby optimizing computational resources and accelerating scientific discovery.

Dflow APIs facilitate the management of workflows and provide real-time status tracking, enhancing the control and transparency of computational processes. Additionally, Dflow includes tools for artifact repository management, enabling efficient upload and download of files. To streamline the development process, it enables the direct upload of local packages into the container's \verb|$PYTHONPATH|,  negating the need for recurrent image builds and easing the integration of user-defined modules. Dflow automatically incorporates its own package into the container by default, satisfying runtime dependencies without necessitating a separate installation in the user's images.

\subsection{Super OP, recursion and conditions}

Steps and DAG are two advanced categories of OP templates which are defined by their constituent steps or tasks instead of a container. The difference between them is that steps are executed consecutively while tasks within a DAG are performed according to their dependencies. Dflow will automatically identify dependencies among tasks within a DAG based on their input/output relationships, with users retaining the option to specify additional dependencies.

A steps/dag can be regarded as a sub-workflow or a super OP template composed of some smaller OPs. Parameters and artifacts can be transferred within a steps/dag just like the case for a workflow. Input parameters/artifacts can be defined for a steps/dag, which can be used as inputs by each step/task within it. One can also declare output parameters/artifacts for a steps/dag and their source (e.g. from outputs of a step/task within it). A steps/dag can be used as template to instantiate a \emph{larger} step just like a container template.

Steps/DAG allows for construct complex workflows of nested structure. It is also possible to recursively use a steps/dag as the template of a building block within itself to achieve dynamic loop.

One can set a step to be conditional so that it will be executed when an expression is evaluated to be true in the runtime, skipped otherwise. It is often used as the breaking condition of a recursive steps.

\subsection{Slices: map and reduce}

In scientific computing, it is often required to generate a list of parallel steps which share a common OP template, and only differ in the inputs, and then collect their results together. E.g., performing the same calculations using several settings in parallel.

Slices helps user to slice input parameters/artifacts (each of which should be a list) to feed parallel steps and stack their output parameters/artifacts into lists following the same pattern. Developers simply need to write an OP that handles a single slice. Both Python OP and super OP (Steps/DAG) are supported to construct a sliced step.

\subsection{Exception handling/fault tolerance}

One can declare exception handling policies before submitting a workflow. Dflow catches \verb|dflow.TransientError| raised by OP. One can set maximum number of retries on transient error. Timeout can be set for a step. When the time consumed exceeds the threshold, a timeout error will be raised. Timeout error can be regarded as fatal error or transient error as required.

When a fatal error is raised or the retries on transient error reaches maximum retries. The step is considered as failed. One can enable the workflow to continue even if a step fails. For parallel steps produced by slices, the workflow can be configured to continue when certain number/ratio of parallel steps succeed.

\subsection{Resubmit/restart mechanism}

A unique key can be assigned to each step for the convenience of locating the step. The key of a step can be considered as a special input parameter whose value may vary according to the context. For instance, the key of a step may depends on the iteration of a dynamic loop. Once a step assigned with a unique key is produced, it can be exactly retrieved via \verb|query_step| by the key.

Workflows often include some computationally expensive steps. Outputs of previously run steps can be reused for submitting a new workflow. This allows, e.g. a failed workflow to be restarted from a specific point after modifying the workflow template or even the outputs of completed steps.

To reuse steps from a previous workflow, first, retrieve the steps to be reused of the previous workflow by \verb|query_step|. If modifications of existing outputs is required, use \verb|modify_output_parameter| or \verb|modify_output_artifact| before reusing. Finally, pass the list of reused steps to the argument \verb|reuse_step| when submitting a new workflow. During the new workflow running, it will detect if there exists a reused step with a matching key. If a match is found, the workflow will skip the step and use the outputs from the reused step.

\subsection{Executor plugins: invoking external computing resources}

By default, an ``executive step'' (a step using a container OP template) executes the shell script or Python script directly within the container. Alternatively, one can specify the executor to run the script. Dflow offers an extension point for executive steps. An executor is an instance of class derived from abstract class \verb|Executor|. An executor should implement a method \verb|render| which transforms original template into a new template. The new template, for instance, submits the script to a job management system or other computational environments for execution. 
The executor can also be designated for a workflow, serving as the default executor that affects every executive step within it. In this case, it is still supported to customize executor for specific steps, overriding the default behavior.
Executors are usually used for invoking different computing resources (out of cluster).

DPDispatcher\footnote{\url{https://github.com/deepmodeling/dpdispatcher}} is a python package used to generate HPC scheduler systems (Slurm/PBS/LSF) or Bohrium jobs input scripts and submit these scripts and poke until they finish. Dflow provides a simple interface \verb|DispatcherExecutor| to utilize DPDispatcher as an executor to accomplish executive steps. One only needs to configure some necessary arguments of \verb|DispatcherExecutor| (including machine and resource parameters in DPDispatcher) before assigning the executor to a step. There are three options for SSH authentication to the login node of the slurm cluster, using password, specifying path of a local private key file, or distributing authorized private keys to each Kubernetes node (or equivalently add each Kubernetes node to the authorized host list).

One can also use virtual node technique to uniformly manage HPC resources within the Kubernetes framework. The wlm-operator\footnote{\url{https://github.com/dptech-corp/wlm-operator}} project is implemented as a Kubernetes operator. Each HPC partition (queue) is represented as a virtual node in Kubernetes with labels representing resource properties of the partition, including CPUs, memory, nodes, wall-time, etc. This allows the Kubernetes to schedule jobs on a suitable partition with enough resources smartly. Within the wlm-operator framework, a virtual Pod instead of an actual Pod is used for monitoring an HPC job, thereby reducing resource costs. Once the wlm-operator is deployed to the Kubernetes cluster, \verb|SlurmJobTemplate| with a node selector can be used to bind a step to one or multiple specified HPC partition(s).

\subsection{Debug bare-metally}

Dflow provides a debug mode for running workflows bare-metally whose backend is implemented in Dflow in pure Python, independent of Argo/Kubernetes. The debug mode utilizes local environment to execute OPs instead of containers as well as local file system to store data by default. It implements most APIs of the default (Argo) mode in order to ensure a consistent user experience. The debug mode offers convenience for debugging or testing without container. In cases where clusters face difficulties deploying Docker or Kubernetes or have limited privilege, the debug mode may also be used for production.

Before running a workflow locally, ensure that the dependencies of all OPs within the workflow are properly configured in the local environment. This is not necessary when using the dispatcher executor to submit jobs to some remote environments. In the debug mode, each workflow will create a new directory locally with a particular structure. The top level of the workflow directory contains the workflow's status and all its steps. The directory name for each step will be its key if provided, or generated from its name otherwise. Each step directory contains the input/output parameters/artifacts, type and phase of the step. For a step of type ``Pod'', its directory also includes the script, log file and working directory for the step.

\subsection{Artifact storage plugins}

The default storage for artifacts in Dflow is a Minio server deployed in the Kubernetes cluster. While it can be seamlessly replaced with various artifact storages (e.g. Aliyun OSS, Azure blob storage (ABS), Google cloud storage(GCS)). Dflow provides an extension point to use customized storage in the artifact management. This can be achieved through a storage client \verb|StorageClient|, a class implementing 5 methods, \verb|upload|, \verb|download|, \verb|list|, \verb|copy| and \verb|get_md5| (optional), which offer the functionality of uploading file, downloading file, listing files with specific prefix, copying file on the server side and getting the MD5 sum of file, respectively.

\section{Applications}

In comparison to traditional workflows, AI-for-science workflows often work with a multitude of independently developed and self-contained modules that operate concurrently. These modules are decoupled yet cohesive, allowing for the flexibility to adapt and scale according to specific AI-for-science needs. This independent development aligns with the scientific community's preference for modular programming, where each component can be developed with writing separate unit test (UT). In practice, these modules often operate with a functional-style approach, standardized around fixed inputs and outputs, a design philosophy that Dflow incorporates in its workflow construction.

AI-for-science workflows frequently utilize diverse and large-scale computational resources, particularly for the development of large atomic models (LAMs)~\cite{zhang2023dpa}. For instance, while the training of complex models can be accelerated through GPU-based parallel processing, CPU-intensive first-principle calculations can be effectively managed through concurrent multi-CPU task execution. Dflow's extensible and scalable architecture, coupled with the consistency introduced by containerization, supports the dynamic allocation of those resources on both cloud and high-performance computers. With its support for modularity, functional-style operations, and efficient resource management, Dflow empowers scientists to develop and execute AI-for-science workflows that are not only scalable but also highly adaptable to the evolving demands of scientific research.

In the sections that follow, we'll highlight Dflow's applications in AI-for-science, ranging from building reusable OPs to deploying complex computational suites. This overview will demonstrate Dflow's capacity to tap into a wide array of AI-for-science tools, underscoring its role as a versatile platform for both utilizing existing software and fostering the development of new AI-for-science solutions.

\subsection{FPOP: A collection of OPs for first-principle calculation} \label{fpop}

The OPs in Dflow are reusable in general, but higher-level reusability still requires further design. For example, most developers create their own OPs for preparing and running VASP tasks from scratch. To address this issue, we need to establish and maintain collections of OPs to achieve higher levels of reusability and standardization. Taking first-principles calculations as an example, we manage a collection of OPs related to DFT calculations. The project is called FPOP\footnote{\url{https://github.com/deepmodeling/fpop}}.

\begin{figure}[htb] 
 \center{\includegraphics[width=0.8\textwidth]{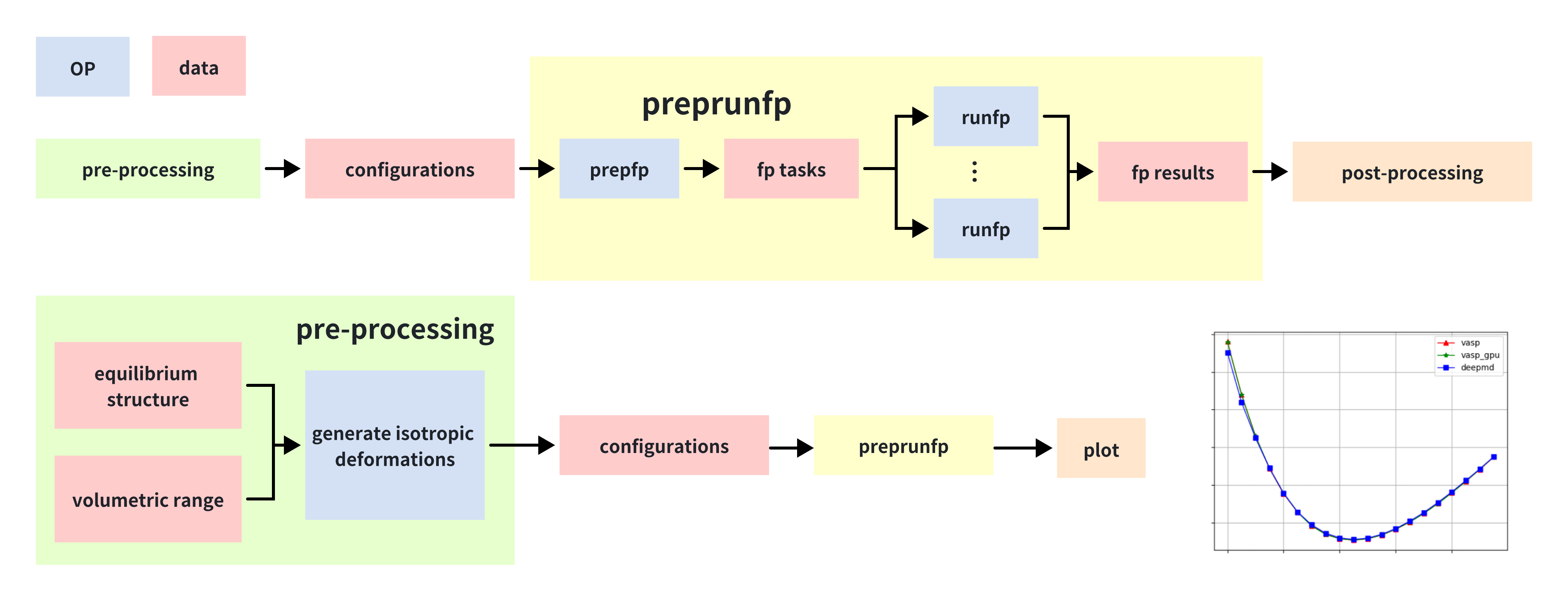}} 
 \caption{\label{fpop-flow} A schematic diagram for an abstract FPOP flow and a flow for calculating equation of state.
 } 
\end{figure}

A typical flow in many DFT calculation tasks is demonstrated in Figure \ref{fpop-flow}. The preprocessing step generates a series of configurations. Then the \texttt{prepfp} step prepares input files for multiple first-principles calculation tasks based on these configurations. Subsequently, the \texttt{runfp} step runs these first-principles calculation tasks concurrently. Finally, the results of the first-principles calculations are collected and post-processed to obtain the final results. 
The preparation and running OPs constitutes a super OP \texttt{preprunfp}.
A specific flow of calculating equation of state (EOS) is depicted in Figure \ref{fpop-flow}. The prepfp and runfp steps exhibit a high degree of reusability, we wrap them into a super OP called preprunfp, which can be directly used to assemble various workflows. Moreover, by abstracting certain input and output formats, FPOP decouples the difference among different DFT calculation softwares and the workflow, making it easy to switch DFT calculation engine in the user's workflow.

To maximize reusability, we carefully designed the parameters exposed to users in FPOP, including necessary inputs (configuration files), optional first-principles calculation parameters (pseudo-potential selection, k-point settings, etc.), runtime environments (image, execution commands, etc.), and workflow logic (fault tolerance, retry logic, etc.). So far, several projects have incorporated FPOP as a component in their workflows, including APEX (Section \ref{apex}) and DPGEN2\footnote{\url{https://github.com/deepmodeling/dpgen2}} (the second-generation DP potential generator).

\subsection{APEX: Alloy Property EXplorer} \label{apex}

\begin{figure}[htb] 
 \center{\includegraphics[width=0.7\textwidth]{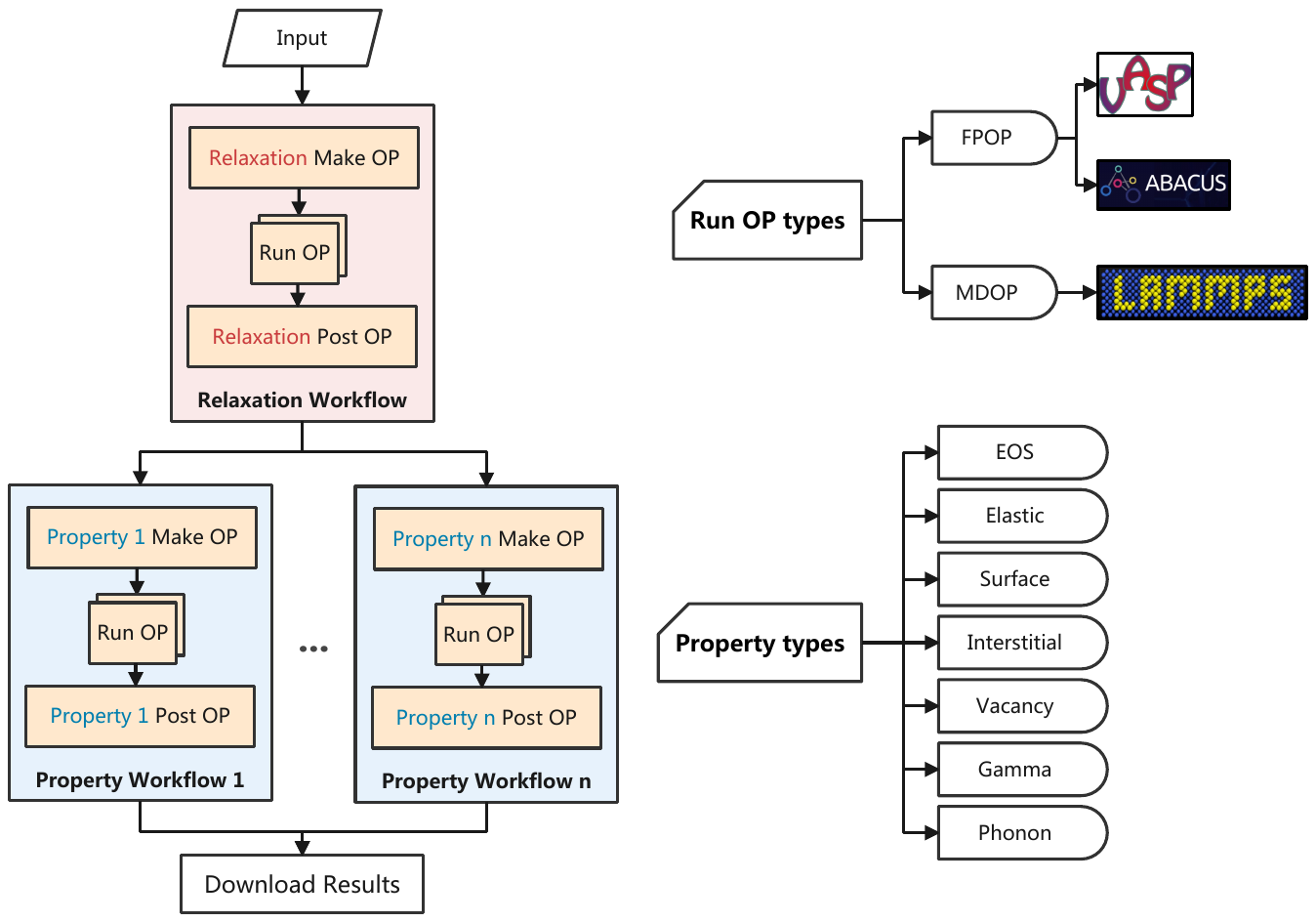}} 
 \caption{\label{APEX-back} The overall architecture of APEX.}
\end{figure}

The APEX\footnote{\url{https://github.com/deepmodeling/apex}} (Alloy Property EXplorer) package is an open-source, extendible, and cloud-native platform for effectively evaluating different interatomic potentials and generating extensive materials property datasets. APEX is built on Dflow and features a modular architecture, as depicted in Fig.~\ref{APEX-back}. It employs Docker for managing different containers in the framework of Dflow, effectively separating computing and scheduling logic. APEX is capable of concurrently running density-functional theory (DFT) calculations using VASP and ABACUS through the FPOP (Section \ref{fpop}), as well as molecular dynamics (MD) simulations using LAMMPS for various structures and properties.

APEX currently has three predefined job types including ``relaxation'', ``property'', and ''joint''. The ``relaxation'' job type optimizes given structures, while the ``property'' job type calculates different properties, including

\begin{itemize}
    \item Equation of state (EOS)
    \item Elastic constants
    \item Surface energy
    \item Interstitial formation energy
    \item Vacancy formation energy
    \item Generalized stacking faulty energy (Gamma line)
    \item Phonon spectra
\end{itemize}

\noindent The ``joint'' workflow combines ``relaxation'' and ``property'' to streamline the process of structure optimization and property calculations. The procedure of either ``relaxation'' or ``property'' job can be divided into three steps: (1) preparation of structures and input files, (2) execution of numerous DFT/MD calculation tasks on CPU/GPU machines, and (3) post-processing of results. The OPs are assembled and integrated by Dflow, which facilitates the efficiency and robustness of APEX. APEX is extendible, enabling users to easily customize and implement new property calculations and simulation tools, by virtue of its object-oriented programming and the design of Dflow OP. Additionally, APEX incorporates user-friendly features such as results visualization and NoSQL database client for data storage. For more information about APEX, please refer to~\cite{li2024extendable}.

\subsection{Rid-kit: Reinforced dynamics}

Rid-kit\footnote{\url{https://github.com/deepmodeling/rid-kit}}, developed atop the Dflow workflow engine, is a software package designed for enhanced sampling through reinforced dynamics (RiD)\cite{zhang2018reinforced,wang2022efficient}. This method, grounded in the principles of collective variable (CV) based strategies, aims to surmount the temporal scale challenges inherent in molecular dynamics (MD) simulations. Such challenges are predominantly due to potential energy barriers that significantly hinder the simulation's ability to reach experimental timescales for complex phenomena, such as protein folding\cite{lindorff2011fast}.

The RiD methodology employs a cyclical process that begins with a biased MD simulation for exploration. Resultant conformations are clustered to identify distinct configurations, which guide subsequent MD simulations to determine the mean force in the labeling step. Neural networks are trained during the training phase to approximate the free energy surface (FES), serving as a bias potential for the next biased MD simulation, thus starting a new cycle.


In practical applications, the RiD methodology faces challenges due to the complexity of its concurrent learning nature, as well as the management of parallel jobs and computational resources. Rid-kit addresses these challenges by streamlining the RiD workflow using Dflow, providing a versatile framework for implementing the RiD processes and seamless integrating with various computational platforms, ultimately enhancing the efficiency and accessibility of advanced molecular dynamics simulations.


\begin{figure}[h!]
    \centering
    \includegraphics[width=0.75\textwidth]{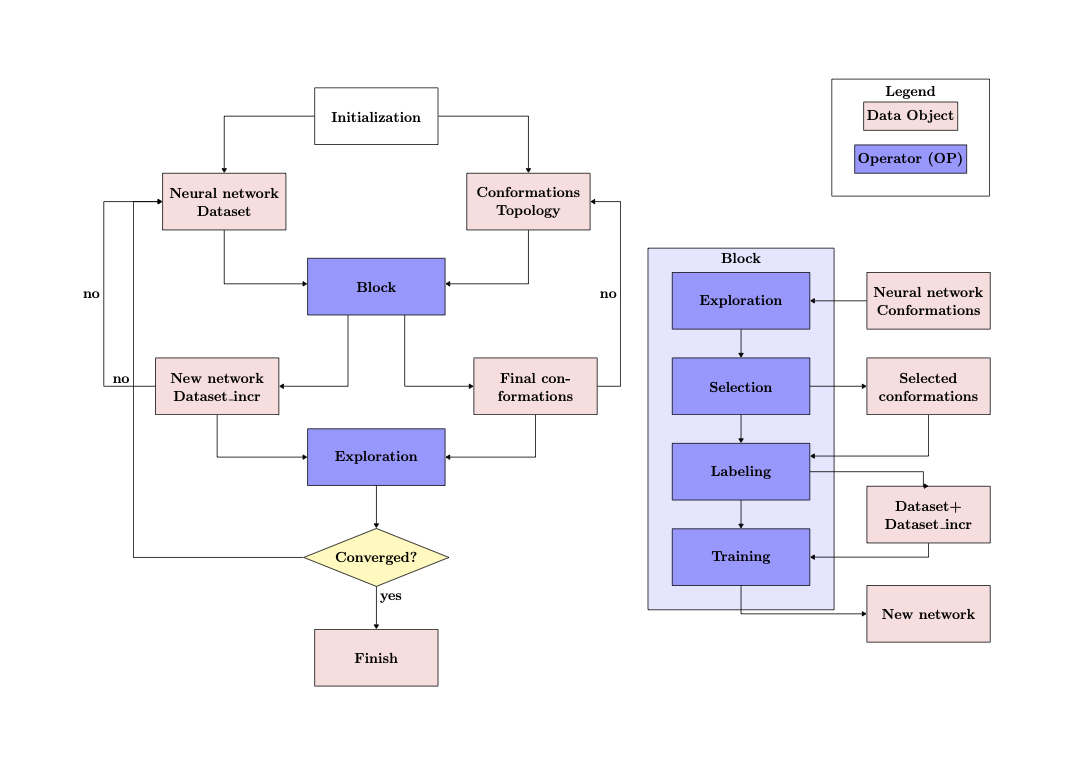}
    \caption{The workflow of Rid-kit.}
    \label{fig:RiD2}
\end{figure}

The computational graph depicted in Figure \ref{fig:RiD2} illustrates the RiD-kit workflow, wherein the core process is executed within a super OP \texttt{Block} during a single iteration. Initially, \texttt{Block} OP receives structure and topological files as inputs for MD simulations. Following each iteration, \texttt{Block} OP updates the neural networks, incorporates new data, and generates new conformations to serve as the initial configurations for MD simulations in the subsequent iteration. Within the \texttt{Block} OP, there are four distinct OPs encompassed, which are as follows

\begin{enumerate}
    \item The \texttt{Exploration} OP uses conformations and neural networks from the previous iteration to perform biased MD simulations, producing new conformations. The implementation of this OP uses Slices to run multiple MD simulations on different GPUs concurrently. Meanwhile, leveraging the Dispatcher executor of Dflow, tasks can be submitted to run on either local multi-GPU machines, HPC environments, or cloud resources.

    \item The \texttt{Selection} OP then takes the conformations generated by \texttt{Exploration} OP, identifies those with high uncertainty for further processing in the \texttt{Labeling} step. Since this step does not consume much computing resources, it can be efficiently accomplished on a 1 or 2-core CPU machine.

    \item The \texttt{Labeling} OP conducts restrained or constrained MD simulations on the conformations selected by \texttt{Selection} OP to calculate mean forces, thereby generating the dataset for training. This step often runs hundreds of independent MD simulations for different conformations, each taking minutes to hours depending on the system. These simulations are well-suited to be done in parallel to speed up the process. In practice, the degree of parallelism can be specified based on the user's requirements (with a default parallelism of 10).

    \item Completing the iteration, the \texttt{Training} OP updates the model by training on the accumulated dataset, resulting in an updated neural network file that represents the free energy model. This is again parallelized by Slices to perform multiple training tasks (default is 4) on different GPUs concurrently.
\end{enumerate}

\subsection{DeePKS flow}
DeePKS \cite{chen2020deepks} is a machine learning-based framework for constructing an accurate energy functional in a self-consistent scheme within the framework of generalized Kohn–Sham density functional theory (DFT). It seeks a Hamiltonian in the generalized Kohn-Sham framework by connecting a computationally efficient baseline method with a highly accurate but costly reference method through a neural network model.

\begin{figure}[htb]
 \centering
 \includegraphics[width=0.4\textwidth]{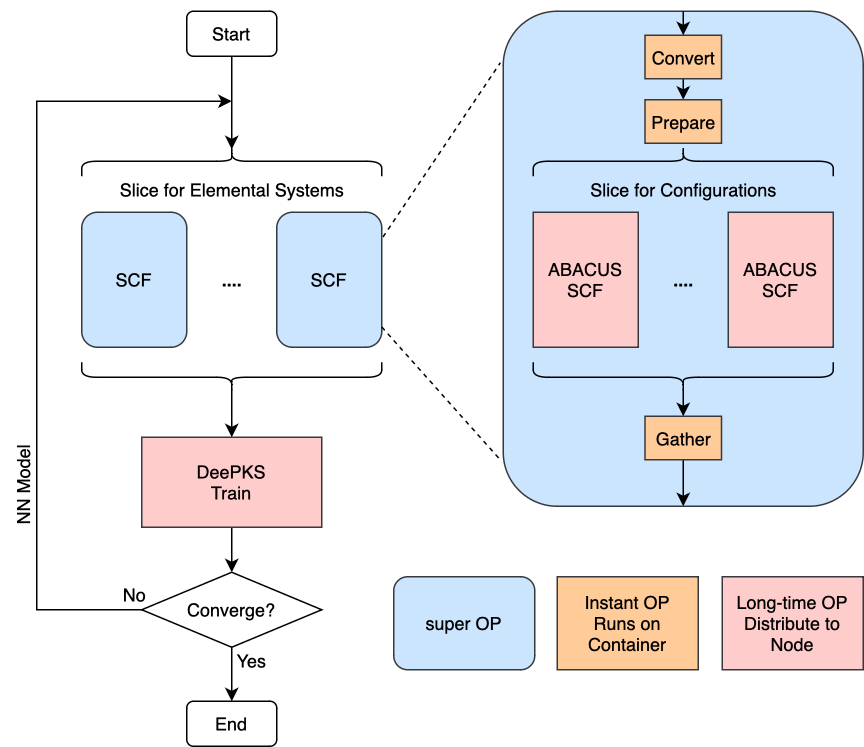}
 \caption{\label{deepks-flow} A schematic diagram for DeePKS flow.}
\end{figure}

The initialization of the flow requires a certain amount of high level Quantum Mechanics (QM) data as the target. As illustrated in Figure \ref{deepks-flow}, the self-consistent iterations in DeePKS consist of two major sections, training the model and solving the Kohn-Sham self-consistent field (SCF) problem

\begin{enumerate}

    \item (SCF) Construct a DeePKS Hamiltonian and solve the generalized Kohn-Sham SCF equations. The DeePKS Hamiltonian is formulated as $\hat{h}_{\mathrm{DeePKS}}=\hat{h}_{\mathrm{baseline}}+\hat{V}^{\delta}$, where $\hat{V}^{\delta}$ denotes the correction potential introduced by the neural network model. This step involves performing independent computations on numerous configurations, which are characterized as CPU-intensive and highly time-consuming processes. The computational efficiency can be significantly enhanced by increasing the degree of parallelism.

    \item (TRAIN) Train the neural network (NN) energy functional. Descriptors in Step 1 serve as inputs for the NN model. The loss function is constructed as the error between the DeePKS results and the reference labels. If the error surpasses the convergence threshold, the iteration continues to Step 1. This step can only be accomplished in a single task. Acceleration of the training process is typically achieved through the utilization of GPU.

\end{enumerate}

The DeePKS flow\footnote{\url{https://github.com/hustlingFive/deepks-flow}}, represents a refactored version of the original DeePKS workflow implemented in DeePKS-kit\footnote{\url{https://github.com/deepmodeling/deepks-kit}}, leveraging the Dflow architecture. In this project, ABACUS is employed to solve the SCF equations.

The utilization of Dflow enhances efficiency by selectively dispatching distinct computational tasks to different resources based on the requirements. The computational performance of SCF is significantly optimized by distributing across numerous CPU machines. Meanwhile, the DeePKS model training process can harness the computational acceleration provided by GPUs. Futhermore, the fault-tolerant feature of Dflow is employed in the DeePKS flow, which allows a certain proportion of SCF calculations to fail without affecting the overall process. Loop-breaking criteria are dynamically determined based on the current iteration.

By virtue of the nested OP feature in Dflow, the SCF OP is constructed as a super OP consisting of smaller OPs for preparation, calculation and post-processing, as illustrated in Figure \ref{deepks-flow}. The DeePKS flow can encapsulate multiple SCF OPs for different elemental systems, thereby support for training a universal DeePKS model for a family of systems.

\subsection{Virtual screening workflow for drug design}

The Virtual Screening Workflow (VSW)\footnote{\url{https://github.com/dp-yuanyn/virtual-screening-workflow}} is an essential component of the Hermite\textsuperscript{\textregistered}\footnote{\url{https://hermite.dp.tech}} Computational Drug Design Platform, which aims to identify potential active compounds from large-scale molecular databases targeting a specific protein target during the early stages of drug design. 
The workflow integrates various tools, including molecular docking, molecular dynamics simulation, free energy calculation, and cheminformatics analysis. 

\begin{figure}
    \centering
    \includegraphics[width=0.98\textwidth]{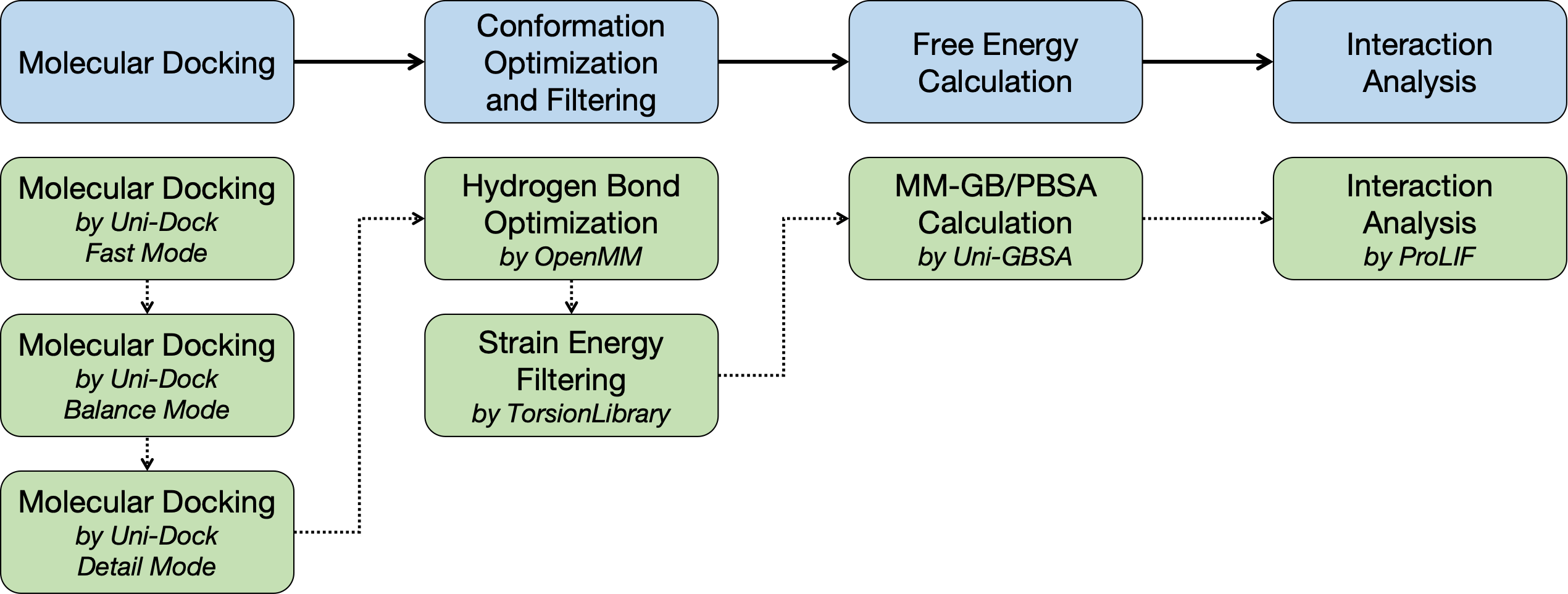}
    \caption{A schematic diagram for VSW.}
    \label{fig:VSW}
\end{figure}

The VSW employs a multi-stage screening strategy~\cite{yu2023uni} for large-scale molecular databases, reducing the actual computational load without significantly affecting the results by designing screening funnels with varying computational complexity and accuracy. As shown in Figure~\ref{fig:VSW}, the workflow consists of the following steps:

\begin{itemize}
    \item Molecular Docking: Uni-Dock~\cite{yu2023uni} is utilized to perform Fast, Balance, and Detail modes of molecular docking. The first round uses the user-input molecular libraries, while the subsequent rounds use the top-ranked results from the previous round as input molecular libraries. After the molecular docking step, a group of potential active molecules with accurate docking score function can be obtained.
    \item Conformation Optimization and Filtering: OpenMM~\cite{eastman2017openmm} is employed for molecular dynamics simulations to optimize the orientations of hydrogen atoms on ligands to form correct hydrogen bonds between ligand and receptor. Subsequently, TorsionLibrary~\cite{penner2022torsion} is used to evaluate and filter out molecular conformations with high strain energy. After the conformation optimization and filtering step, a group of molecules with good protein-ligand binding poses and low ligand strain energy can be obtained.
    \item Free Energy Calculation: Uni-GBSA~\cite{yang2023uni} is used for MM-GB/PBSA calculations to evaluate the binding free energy based on the receptor-ligand binding poses provided by the previous step. The calculated binding free energy is then used for ranking and screening. After the free energy calculation step, a group of potential active molecules with MM-GB/PBSA accuracy can be obtained.
    \item Interaction Analysis: ProLIF~\cite{bouysset2021prolif} is employed to perform interaction type and statistics between receptor-ligand binding poses, providing auxiliary decision-making information for users.
\end{itemize}

The complexity inherent in computational processes presents challenges when constructing VSW. The diversity of software environments required for each step, if consolidated into a single environment, could spark numerous conflicts. Furthermore, the computational demands fluctuate between GPU and CPU, making the allocation of static computational resources inefficient. When VSW is tasked with sifting through libraries housing tens of millions of molecules, the necessity of elastic scaling of computational resources on a large scale becomes apparent. These practical considerations amplify the intricacy of workflow construction.


The advent of Dflow has been a boon to the construction of VSW. We have crafted each computational step as a distinct OP, each operating within its own Docker image. This encapsulation renders each runtime environment more streamlined and agile. Meanwhile, the independence of each OP in resource allocation allows for a judicious use of CPU and GPU resources.


In the realm of molecular docking at a scale of tens of millions, Dflow's Slices feature is harnessed to segment the molecular library into manageable units, with each node handling approximately 18,000 molecules. This strategic partitioning ensures that the computational load on any given node is completed within a half-hour window, facilitating the rapid processing of vast molecular libraries. The library is thus distributed across over 600 nodes operating in unison, a parallelism that is within the reach of our available cloud resources. Given the software's requirements to store each molecule in individual files, the potential for disk performance bottlenecks or inode exhaustion is also mitigated by distributing the disk load across multiple machines, thereby enhancing the robustness of VSW.



Within each Slice, serial logic is orchestrated to enable tasks to operate autonomously. This approach circumvents the bottleneck effect, ensuring that the overall computation speed is not constrained by the slowest processes. During large-scale molecular docking, if a minority of molecules lead to computational failures, Dflow's restart mechanism allows us to selectively address and recompute the problematic molecules without redundantly reprocessing successful nodes. Moreover, when confronted with even larger datasets where the integrity of the overall outcome is not critically dependent on every single data point, Dflow's \textit{continue\_on\_success\_ratio} parameter is leveraged for fault tolerance, enabling the VSW to continue operating despite partial failure.


The tangible advantages conferred by Dflow have markedly elevated the efficiency of constructing and deploying large-scale VSW, without introducing a myriad of engineering complexities. Presently, these workflows are effectively operational on the Hermite\textsuperscript{\textregistered} drug computational design platform, where they are entrusted with the daily task of screening libraries comprising tens of millions of molecules. A quintessential workflow encompasses approximately 1,500 OPs and is capable of achieving a maximum concurrency level of over 1,200 GPU computing nodes, testament to the robustness and scalability of our approach.

\subsection{Dflow-galaxy: more possibilities about Dflow}

Dflow-galaxy\footnote{\url{https://github.com/chenggroup/dflow-galaxy}} is a collection of workflows and tools built on top of Dflow and ai\textsuperscript{2}-kit\footnote{\url{https://github.com/chenggroup/ai2-kit}}. By redesigning the data structure and implementation of OPs, it provides users with an easy way to build their workflows.

Dflow-galaxy introduces a \verb|DFlowBuilder| to wrap \verb|ScriptOpTemplate| and implement a set of \verb|Artifact|s, enabling fan-in/fan-out operations. It also serves as a wrapper, making \verb|OutputArtifact| an input of the workflow. This allows operations to pass a pointer to the output target into the operation, making it more straightforward for users. As a result, \verb|BashStep| and \verb|PythonStep| are supported.

\begin{figure}[htb!]
    \centering
    \includegraphics[width=0.98\textwidth]{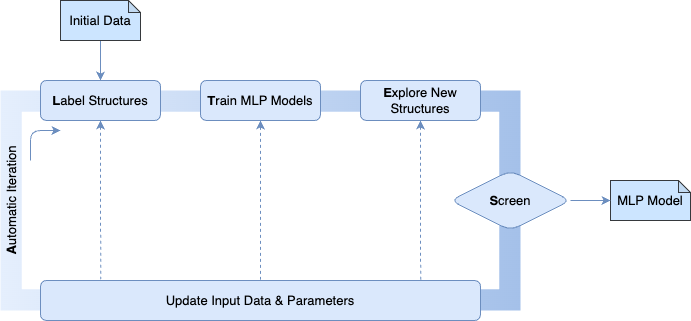}
    \caption{A schematic diagram of the TESLA workflow.}
    \label{fig:TESLA}
\end{figure}

The TESLA workflow (\textbf{T}raining \textbf{E}xploration \textbf{S}creening \textbf{L}abeling \textbf{A}utomatic Workflow) \cite{guoChecMatEWorkflowPackage2023} has been implemented with dflow-galaxy, drawing inspiration from DP-GEN \cite{zhang2020dp} and DP-GEN2. It supports the following steps:
\begin{enumerate}
    \item Train: Train multiple DP potentials with a collected training dataset, creating a model ensemble to estimate the uncertainty of prediction for each frame of the structure.
    \item Explore: Generate Lammps tasks with a given set of parameters using the trained machine learning potentials to explore more structures and expand the training dataset. Special methods like metadynamics run with Plumed \cite{gongAnomalousEntropicEffect2023} or FEP-TI method \cite{wangAutomatedWorkflowComputation2022} are also supported.
    \item Screen: Screen the structures generated by the exploration step based on their model deviation, selecting only those within an acceptable range.
    \item Label: Pass the screened structures for labeling using DFT single-point calculations.
\end{enumerate}
The four steps are combined to create a comprehensive workflow using Dflow. This configuration forms a loop, shown in Figure \ref{fig:TESLA}, enabling iterative exploration of structures that have not yet been accurately predicted in each iteration. This iterative process ensures the MLP is fully trained to effectively study the system.

Compared to the original DP-GEN procedure, dflow-galaxy supports more flexible input of datasets and Lammps input, enabling complex concurrent learning workflows such as the calculation of redox potential \cite{wangAutomatedWorkflowComputation2022}. Powered by Dflow, the orchestration of the workflow can be expanded to different platforms.
With the assistance of the Bohrium platform, \verb|Dynacat TESLA|\footnote{\url{https://app.bohrium.dp.tech/dynacat-tesla/}} has been released on Bohrium Apps alongside \verb|CP2K Lightning|\footnote{\url{https://app.bohrium.dp.tech/cp2k-lightning/}} and \verb|Dynacat MD|\footnote{\url{https://app.bohrium.dp.tech/dynacat-md/}}. These tools enable the training of a suitable MLP and the execution of enhanced sampling MD for dynamic catalysis research from the ground up.

\section{Conclusion}

The advent of the AI-for-science era has brought forth the demands for a new generation of infrastructure in scientific computing scenes such as concurrent learning and high-throughput computing. Dflow, an open-source Python toolkit, meets these demands by providing a user-friendly workflow framework, bridging the gap between algorithm conception and practical application. With the capabilities of complex process control and task scheduling, Dflow is flexible and adaptable to various environments, efficiently utilizing cloud resources and high-performance computing.
While containers simplify environment provisioning and enhance reproducibility, Dflow additionally offers the ability of local debugging without containers. These features have made it the foundation for numerous workflow projects across diverse scientific fields. Moreover, the growth of the Dflow ecosystem and its emphasis on collaboration and extendibility have ensured that it remains at the forefront of workflow management solutions, empowering researchers and developers to drive innovation and accelerate scientific discovery in this dynamic era of AI for science.

\section*{Acknowledgements}

We would like to thank Jiankun Pu for his substantial contributions to the development of Dflow, including his efforts in supplementing the documentation and promoting the project externally.
The work of H.~W.~is supported by the National Key R\&D Program of China (Grant No.~2022YFA1004300) and the National Natural Science Foundation of China (Grant No.~12122103). 
D.M.Y. is grateful for financial support provided by the National Science Foundation (CSSI Frameworks Grant no. 2209718).

\section*{Competing interests} 
The authors declare no competing interests.

\bibliographystyle{johd}
\bibliography{bib}

\end{document}